%% file: main.tex
\PassOptionsToPackage{table}{xcolor}

\documentclass[lettersize,journal]{IEEEtran}

\usepackage[hidelinks]{hyperref}
\usepackage{url}

\usepackage[english]{babel}
\usepackage{blindtext}

\usepackage{float}
\usepackage[utf8]{inputenc}
\usepackage{hyperref}
\usepackage{url}

\usepackage{fancyhdr}

\usepackage{graphicx}
\usepackage{xspace}
\usepackage{tcolorbox}

\usepackage{pdflscape}
\usepackage{multirow}
\usepackage{amsmath}
\usepackage{tabto}
\usepackage{array}
\usepackage{indentfirst}
\usepackage{lscape}
\usepackage{adjustbox}
\usepackage{pgfgantt}
\usepackage{eurosym}

\usepackage{listings}
\usepackage{caption}

\usepackage{soul}

\usepackage{titlesec}

\usepackage[table]{xcolor}
\definecolor{azulUC3M}{RGB}{0,0,102}
\definecolor{gray97}{gray}{.97}
\definecolor{gray75}{gray}{.75}
\definecolor{gray45}{gray}{.45}
\usepackage{multicol}
\usepackage{paralist}

\definecolor{codegreen}{rgb}{0,0.6,0}
\definecolor{codegray}{rgb}{0.5,0.5,0.5}
\definecolor{codepurple}{rgb}{0.58,0,0.82}
\definecolor{backcolour}{rgb}{0.95,0.95,0.92}
\definecolor{mygreen}{rgb}{0,0.6,0}
\definecolor{mygray}{rgb}{0.5,0.5,0.5}
\definecolor{mymauve}{rgb}{0.58,0,0.82}

\newcommand\extended[1]{{}} 

\newif\ifstatus
\statusfalse 
\sethlcolor{white} 

\newcommand\alfonso[1]{\ifstatus\textcolor{black!50!red}{#1 - Alfonso}\fi}

\newcommand\guille[1]{\ifstatus\textcolor{black!50!blue}{#1 - Guille}\fi}

\newcommand\future[1]{\textcolor{black!50!gray}{}}

\newcommand\todolist{{\sc{TodoList}}}

\usepackage{dirtree}

\begin{document}

\title{

Snorkeling in dark waters: A longitudinal surface exploration of unique Tor Hidden Services (Extended Version) 

\thanks{
A shorter version of this paper is published in IEEE Transactions on Information Forensics and Security. This is the extended version.\\
\copyright~ 2025 IEEE. Personal use of this material is permitted. 
Permission from IEEE must be obtained for all other uses, including reprinting/republishing this material for advertising or promotional purposes,  creating new collective works, for resale or redistribution to servers or lists, 
or reuse of any copyrighted component of this work in other works.
}

}

\author{Alfonso Rodriguez Barredo-Valenzuela\textsuperscript{\dag}\textsuperscript{\ddag}, Sergio Pastrana Portillo\textsuperscript{\ddag}, Guillermo Suarez-Tangil\textsuperscript{\dag}\\
\textsuperscript{\dag}\textit{IMDEA Networks Institute}, \textsuperscript{\ddag}\textit{Universidad Carlos III de Madrid}}

\maketitle

\begin{abstract}
    The Onion Router (Tor) is a controversial network whose utility is constantly under scrutiny. 
    On the one hand, it allows for anonymous interaction and cooperation of users seeking untraceable navigation on the Internet. 
    This freedom also attracts criminals who aim to thwart law enforcement investigations, e.g., trading illegal products or services such as drugs or weapons. 
    Tor allows delivering content without revealing the actual hosting address, by means of \texttt{.onion} (or hidden) services. 
    Different from regular domains, these services can not be resolved by traditional name services, are not indexed by regular search engines, and they frequently change. 
    This generates uncertainty about the extent and size of the Tor network and the type of content offered. 
    
    In this work, we present a large-scale analysis of the Tor Network. We leverage our crawler, dubbed Mimir, which automatically collects and visits content linked within the pages to collect a dataset of pages from more than 25k sites. We analyze the topology of the Tor Network, including its depth and reachability from the surface web. 
    We define a set of heuristics to detect the presence of replicated content (mirrors) and show that most of the analyzed content in the Dark Web ($\approx$82\%) is a replica of other content. 
    Also, we train a custom Machine Learning classifier to understand the type of content the hidden services offer. 
    \hl{Overall, our study provides new insights into the Tor network, highlighting the importance of initial seeding for focus on specific topics, and optimize the crawling process. We show that previous work on large-scale Tor measurements does not consider the presence of mirrors, which biases their understanding of the Dark Web topology and the distribution of content.}
\end{abstract}


\input{sections/introduction}

\input{sections/methodology}
\input{sections/analysis}
\input{sections/use_case}
\input{sections/discussionLimitations}

\input{sections/relatedWork}
\input{sections/conclusions}


\bibliographystyle{IEEEtranS}
\bibliography{bibliography}

\end{document}

%% file: sections/introduction.tex
\section{Introduction}

The Onion Router (Tor) has grown considerably since its emergence in 2002~\cite{kadianakis2015extrapolating,bernaschi2022onion,zabihimayvan2019broad}, 

and has become a widely used platform for users and communities seeking privacy and anonymity while navigating the Internet. For example, it is the main social technology used for cooperative work for marginal communities, such as LGTBI+, in countries where their activities are forbidden and even penalized~\cite{devito2019socialtechnologies}. Moreover, Tor offers a preferred online space for political activists to organize and share information~\cite{forte2017privacy}. Besides allowing clients to navigate anonymously, Tor also allows content providers to offer web services without revealing the actual IP address hosting the server. This is done using so-called \textit{hidden services} or \textit{.onion sites}. To access these services, users need to establish a connection through the Tor Network, thus preserving the anonymity of clients and servers. This neutral technology poses a dilemma since it can be used both for benign purposes (e.g., censorship avoidance, whistle-blowing, and activism), and all sorts of \hl{malicious} services (e.g., drug dealing, terrorism,, distribution of child abuse material, or data breaches)~\cite{jardine2015dark}. 

The Tor Network provides the infrastructure to what is referred to as the Dark Web, i.e., the portion of the Web that can not be easily accessed by standard tools. 

The Dark Web often hosts marketplaces trading digital goods that are key to conducting cyberattacks~\cite{van2018plug}. 
The Dark Web also hosts online forums where criminals often publish data leaks from ransomware attacks or data breaches~\cite{breachsense}. 
Consequently, the analysis of the Dark Web is important to understand modern cyber threats and adversaries.
However, there are several challenges that analysts must face when collecting and analyzing such information, like the volatility of the information or the unreliability of the connections.

Various research works have studied the Dark Web in the past, most of them focusing on specific content (e.g., marketplaces~\cite{soska2015measuring,van2018plug,labrador2022examining} or criminal activities~\cite{kloess2021trust,finklea2017})

These studies help to understand a niche of the Dark Web but do not provide a general overview of its size, connectivity, and overall contents offered. 
Prior work that provides an overview of the Tor network dates back several years~\cite{owenson2015tor,duta10k-al2019torank,faizan2019exploring,avarikioti2018structure,bernaschi2022onion,bernaschi2019spiders} and does not consider the prevalence of replicated hidden services scattered through the Tor network.
Thus, they offer a biased and outdated picture of the prevalence of content on the Dark Web. 
As Tor is a volatile network~\cite{bernaschi2022onion,Burda19} with duplicated content across the board, there is a need for a more comprehensive and generic study of the topology and the prevalence of \hl{different types of} contents on the Dark Web.

In this work, we propose a crawling methodology designed to longitudinally and reliably map second-level domain services in Tor. 
We conduct shallow, but in-breath, crawling, i.e., once the crawler visits the landing page of a site, it moves to different ones --- avoiding an in-depth exploration of the site.
Our aim is to maximize the crawling coverage of the Dark Web to capture \hl{sites related to deviant content (i.e., ``actions that violate social norms, which may include both informal social rules or more formal societal expectations and laws''}~\cite{deviant}\hl{), often linked to cybercrime activities}, without overloading the network with a deeper crawling of the sites we visit. As such, sites of particular interest can be efficiently detected to conduct in-depth analysis (as we show in the case study described in \S\ref{sec:use_case}). 
As a key distinction, our methodology pinpoints Web pages that are predominantly the same or very similar (namely mirrors). 
We thus perform the first mirrorless network analysis to study the topology and the type of content hosted on the Dark Web, their language, and the role that mirrors play when measuring hidden services behind Tor.
We present a case study of how our crawler reaches 159 sites hosting child abuse content, including live cams with pornographic content for pedophiles. 

Overall, our main contributions are:
\begin{itemize}
    \item We build a custom crawler (Mimir) that allows us to efficiently navigate through the Tor Network (\S\ref{sec:methodology}). 
    The crawler departs from a set of seeds that are obtained systematically, by querying, in a novel way, dedicated .onion search engines for trending topics in underground surface forums. 
    Our results (\S\ref{sec:analysis}) show that our crawling methods are more effective than those in other works (\S\ref{sec:related}), reaching a larger portion of the network from a reduced set of seeds.

    \item To measure the prevalence of replication, 
    we develop and validate an effective heuristic-based algorithm for mirror detection (\S\ref{sec:methodology}). 
    {We notice that most ($\approx$82\%) of the landing pages in \hl{our dataset} are a replica} (\S\ref{sec:analysis}). 
    This insight is key when accurately measuring the prevalence of genuine \hl{deviant} content (\S\ref{discussions}) in the Dark Web, which we conduct in this work by filtering out duplicates for the first time in the literature.

    \item Our insight above motivates us to characterize the type and study the amount of \hl{activities potentially linked to cybercrime} on the Dark Web. 
    We source a known dataset of labeled sites to build a custom content classifier (\S\ref{sec:methodology}). 
    We design an \hl{efficient} learning pipeline that performs well despite the scarcity of data\extended{, especially in some categories} \hl{. The classification allows us to validate our crawling strategy}.
    We show that a large proportion of \hl{the unique sites of our dataset} (74.4\%) is dedicated to cybercrime activities (\S\ref{sec:analysis}).

   \item As a case study, we illustrate how Mimir detects 159\extended{ (61 still available)} Tor sites related to the distribution of child abuse material, including a site offering live cams over underage (\S\ref{sec:use_case}). These sites have been reported to our Law Enforcement agency and are currently under investigation.
\end{itemize}

\extended{The rest of the paper is organized as follows. First, \S\ref{sec:methodology} describes the methodology used to conduct the study. Second, \S\ref{sec:analysis} describes the experimental setup and analyzes the results of our measurement. Third, \S\ref{sec:use_case} describes the use case for the identification of sites related to child abuse material. Fourth, \S\ref{discussions} describes the limitations and discuss the key findings. And finally, \S\ref{sec:conclusions} presents the conclusions.}

%% file: sections/methodology.tex
\section{Methodology}
\label{sec:methodology}

\begin{figure*}[t]
	\centering
	\includegraphics[width=.9\textwidth, clip]{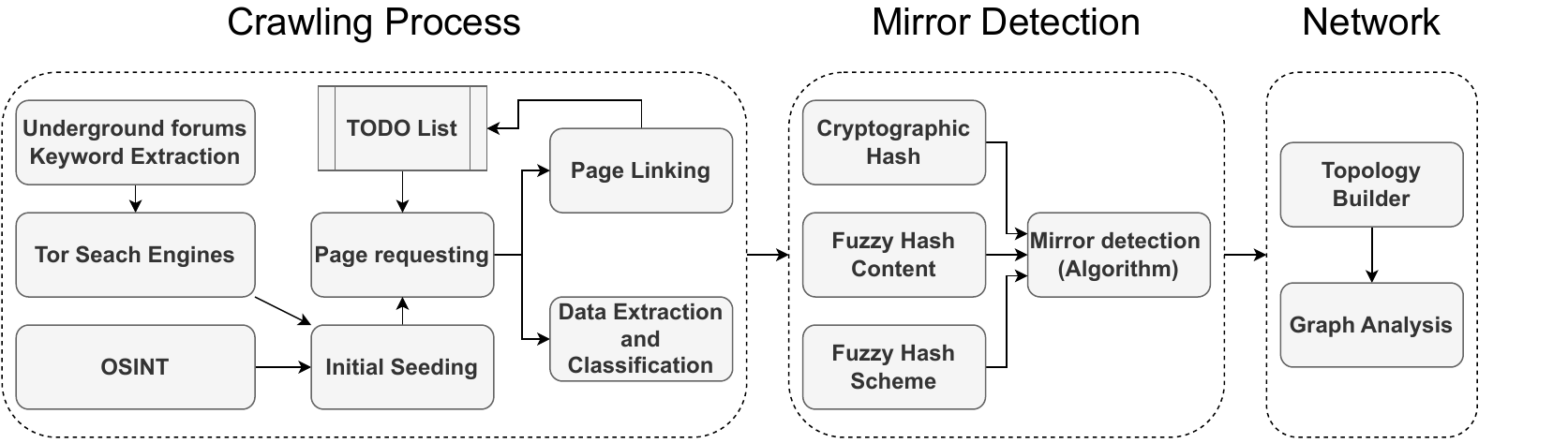}
	\caption{Overview of the methodology followed for our study.
	}
	\label{fig:methodology}
\end{figure*}

Our methodology comprises three building blocks (see Figure~\ref{fig:methodology}): a crawler for the data collection, a mirror analysis algorithm for the identification of replicated (or slightly modified) sites, and a module for network analysis and content classification \hl{that serves to validate the effectiveness of our proposed crawling pipeline}. 
It takes as input an initial set of seeds which, unlike prior work, are extracted systematically.

\subsection{Challenges}
Analyzing and comprehensively understanding the Dark Web presents staggering challenges that have an impact on the design of our methodology, which we describe next.

\noindent\textbf{Scheduling strategy.}
Volatility is a well-known characteristic of the Darknet Tor~\cite{bernaschi2022onion}. 
Some Hidden Services (HS) may be up for intermittent periods, and thus it is necessary to attempt connection to a given site multiple times before discarding the link for not being online. 
This increases the overload of the network. 
To be compliant with the safety guidelines of the Tor Research Safety Goals~\cite{safetyboard}, i.e., minimization, we need to find a trade-off between the coverage of our study, and the overload caused by the network. 
Accordingly, one design requirement of our system is to perform a crawling that aims to maximize the number of unique websites encountered, even at the cost of not exploring each site in its entirety. 
As such, our system prioritizes breadth over depth in its crawling strategy and performs what we call a `shallow' crawl. 
By prioritizing breadth, our system gains a wider perspective of the Dark Web.

\noindent\textbf{Content analysis.}
Media content such as images and GIFs are common ways to deliver information. 
Images and GIFs are a common way to convey information. Such media contains valuable information, and it is frequently used in information retrieval processes. 
However, due to legal and ethical advice, we can not use this information in Tor. 
As our case study demonstrates, longitudinal crawling often encounters sites distributing illegal content like child abuse media. 
Note that the download of child abuse content alone is illegal and considered a serious crime. 
Thus, we require our crawler to retrieve only text and our classification systems to perform accurately using non-media content. Other studies such as the one conducted by Claudia Peersman et al. \cite{CSAIMG} have addressed the challenge of analyzing content in child abuse images through a strict collaboration with law enforcement and under a very specific context. 
The broad nature of our crawling process does not meet the necessary requirements. 

\subsection{Crawling Process}
\label{sec:method:crawling}

The crawling process collects (Dark) Web pages using several steps. 
The process starts with the crawling of the sites obtained from an initial set of seeds (which we describe below). 
For each site, the crawler extracts the information defined in Table \ref{data_extracted}, including links to other pages. 
These are added in a \todolist{} for subsequent crawling iterations. 
After a link is successfully crawled, Mimir continues with the next item in the \todolist{} until it is empty.
Since sites may sometimes be offline, and also to ensure that non-availability is not due to an eventual network error, we revisit (at most 5 times) each URL responding with an HTTP code different than 200. 

If an URL remains unavailable after all 5 attempts, we remove it from the \todolist{} and flag it as \textit{Unreachable}.
\extended{Our crawler runs on the background of an online system that automatically synchronizes the URLs it discovers and their attributes with a repository on a regular basis. 
All this using concurrency and a windowed algorithm for dynamic work allocation of the \todolist{}.} 
Next, we detail the crawling process.

\noindent\textbf{Initial seeding.}

We use a systematic approach to collect the initial seed of .onion sites. 
Specifically, we automatically query three popular Tor Search Engines (TSE) for hidden services with relevant keywords: \textit{Ahmia}~\cite{Ahmia}, \textit{Torch}~\cite{Torch} and \textit{VisiTor}~\cite{VisiTor}. 
To provide these keywords, we extract the most relevant words from the forum titles in underground forums~\cite{CrimeBB} using the Term Frequency -- Inverse Document Frequency (TFIDF) approach, as we further explained in Section \ref{subsec:methodContent}. 
Our hypothesis is that discussions in underground forums are a useful source for the initial seeding of \hl{deviant} content in Tor.
\hl{We note that, while this work focuses on content related to deviant behavior, Mimir can be used to investigate other topics by using a different corpus to extract keywords for the seeding process.}
\extended{Appendix~\ref{appendix:words} lists the complete set of keywords, which represent common topics found in underground hacking forums, including references to video game hacks and cheats.}

Finally, to verify that prominent .onion sites appear in our initial seed we use:
\begin{enumerate}
    \item Domain knowledge and references from previous works (e.g., Hidden Wiki or specific GitHub repositories);
    \item Literature search to improve the previous domain knowledge and contextualize the current state of art seeking for new relevant seeds; 
    \item Open Source Intelligence (OSINT), i.e., TSE to collect new seeds or to assess if existing seeds are informative.   
\end{enumerate}
\noindent All of the tasks mentioned above require some level of manual effort but result in the inclusion of additional resources that complement the TSE search.
\begin{table}
	\centering
    \caption{Attributes extracted for each page crawled.}
	\begin{adjustbox}{width=\columnwidth}
        \begin{tabular}{llll}
        	\textbf{Attribute} & \multicolumn{3}{l}{\textbf{Description}} \\
        	\hline
        	\rowcolor[HTML]{EFEFEF} 
        	URL & \multicolumn{3}{l}{\cellcolor[HTML]{EFEFEF}Full \texttt{.onion} address} \\
        	Metadata & \multicolumn{3}{l}{Metadata of the page} \\
        	\rowcolor[HTML]{EFEFEF} 
        	Link list & \multicolumn{3}{l}{\cellcolor[HTML]{EFEFEF}A list of the links contained in the page} \\
        	Referenced & \multicolumn{3}{l}{List of the pages which has references to this page} \\
        	\rowcolor[HTML]{EFEFEF} 
        	HTML & \multicolumn{3}{l}{\cellcolor[HTML]{EFEFEF}The raw HTML code of the page} \\
        	Timestamp & \multicolumn{3}{l}{String with the Timestamp in format ``dd-MM-yy HH:mm:ss"} \\
        	\rowcolor[HTML]{EFEFEF} 
        	Language List & \multicolumn{3}{l}{\cellcolor[HTML]{EFEFEF}A list of the languages contained in the page} \\
        	Depth & \multicolumn{3}{l}{The number of links from the seed to this page} \\
        \end{tabular}
	\end{adjustbox}
	\label{data_extracted}

\end{table}

\begin{figure}[ht]
	\centering
	\includegraphics[width=.99\columnwidth, trim=0 0 0 12, clip]{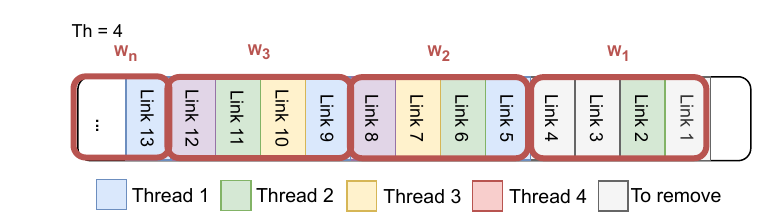}
	\caption{Dynamic workload algorithm example with 4 threads.}
	\label{fig:workload_windows}
\end{figure}

\noindent\textbf{Page requesting.}
\extended{As Mimir is designed to be scalable, we design a concurrent process to request more than one site at the same time. To avoid excessive use of synchronization mechanisms such as mutexes, events, or semaphores, we perform a dynamic work allocation to define which thread will be in charge of requesting each site. It is necessary to decide this dynamically given that the links to be crawled (workload) are stored in the \todolist{} which is a queue with dynamic size. In other words, the number of links is constantly increasing and decreasing according to new URLs that are being added, or sites being removed once they are successfully collected. For this purpose, }
We use \textit{workload windows} to run parallel crawling and dynamically allocate links that need to be crawled by different running threads. 
This strategy avoids the use of additional synchronization mechanisms. 
As depicted in Figure \ref{fig:workload_windows}, Each thread has only one link to crawl per window, as the number of links in each chunk is equal to the number of threads. 
Thus, if a given thread finishes with its link, it crawls $Link_{k+\#threads}$, where \textit{k} is the link number. 
To remove links from \todolist(), all links for the window must be completed (i.e., either a valid HTTP response was received or the request timed out).
Unsuccessful links are added again to the \todolist{} if they have been attempted less than 5 times. Otherwise, they are marked as \textit{unreachable}.
If the links are successfully requested, we run the steps detailed in the following sections.

\noindent\textbf{Data extraction.} 
\label{sec:data-extraction}
We process all .onion sites we crawl and we extract the information shown in Table~\ref{data_extracted}. 
As we crawl, we scrape the list of links to other hidden services and add them to the \todolist{}. 
We also scrape the following information, which is then used for the analysis: 
i) the language of the site, by using {\em langdetect} library \cite{nakatani2010langdetect}, ii) the depth from the seed which leads to a page being currently crawled, and iii) the timestamp of the crawl. 
Finally, we also store the raw HTML, for further content and mirror analysis.\extended{ To scrape and store the HTML, we rely on the Beautiful Soup Python library~\cite{richardson2007beautiful}.}

\noindent\textbf{Page linking}. 
This step uses regular expressions to extract linked .onion pages in a given site, to further guide the crawler.
The relationship between pages lets us determine the network topology of Tor. 
We classify the links we extract into three categories:
\begin{enumerate}
        \item {\em Own links:} Sub-pages hosted within the same domain. 
        \item {\em External \texttt{.onion} links:} These are links to other hidden services {(i.e., pointing to an external \texttt{.onion} site)} extracted from the raw HTML.
        \item {\em Surface links:} These are the links that have a Top Level Domain other than \texttt{.onion}.
\end{enumerate}

Since our goal is to conduct horizontal crawling, only external \texttt{.onion} links are added to the \todolist{} (after removing all characters at the right side of the domain), for being crawled in subsequent iterations. 
During this process, we build a link list for each site that allows us to rebuild the crawling path from the original seed. 
The path is updated to build a network graph upon observing new cross-links. 

\subsection{Mirror Analysis}

To reduce our dataset to a set of singular pages, this step identifies mirrors. A mirror is an exact or nearly equal copy of another \texttt{.onion} site.
Our rationale is that the presence of mirrors introduces bias to existing Tor content- or topology-driven analysis. 
For example, the presence of the same market under different \texttt{.onion} addresses leads to the false impression that there are more markets of the same kind.  
{While most of the mirrors are an exact clone of a site, an exploratory analysis shows a frequent deployment of sites with minor modifications (e.g., a market with the same products but with prices in a different currency --- typically cryptocurrencies vs. dollars). 
We also see the same site translated into different languages and we thus consider them as mirrors.} 
This insight informs the development of a similarity measure robust in the face of inconsequential modifications as we discuss next.

\noindent{\textbf{Web content vs web structure}.}
There are two main elements of a web page that we compare to understand how similar they are: 1) the content of the page, and 2) its structure (namely HTML scheme). 
Comparing two pages based on their content alone would not capture situations where the page offers the same content but in different formats (e.g., translations of the same content into different languages, or pricing in different currencies based on the locale of the user). 
Conversely, relying only on the HTML scheme would flag as mirrors sites that use the same underlying framework (e.g., two blogs with the same WordPress template). 
The scheme is extracted by using an algorithm that keeps only HTML tags.

Our methodology distinguishes this casuistry and applies a custom heuristic-based system that we describe next.

\begin{figure}[ht]
	\centering
	\includegraphics[width=.99\columnwidth]{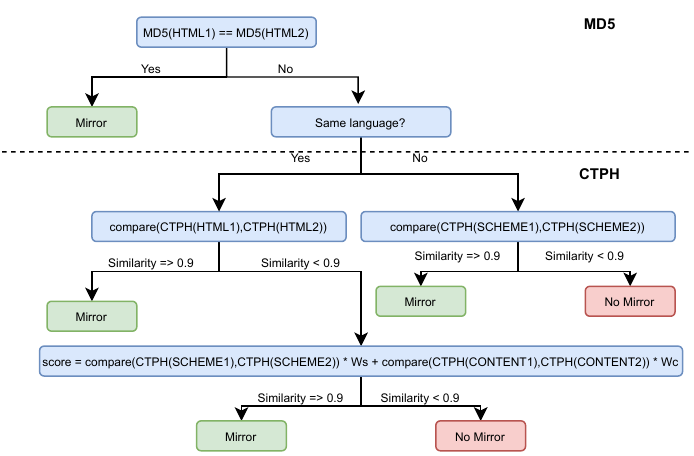}
	\caption{Workflow of the mirror detection algorithm. The output provides a similarity score (1 being the most similar).}
	\label{fig:mirror}

\end{figure}

\noindent{\textbf{Hybrid hashing}.}
To assess if two pages are mirrors, we apply a combined approach, using two different hash functions, as depicted in Figure~\ref{fig:mirror}.  
We first rely on the \textit{cryptographic hashing} algorithm MD5 \cite{rivest1992md5} to determine whether two pages are \textit{exact} copies. 
This has a binary outcome, i.e., the pages are equal or not. 
In case it is not equal, we use a \textit{fuzzy hashing} function that scores how \textit{similar} two pages are, with a certain tolerance for changes. 
Our implementation uses the CTPH algorithm \cite{kornblum2006identifying}, which breaks down pages into several components and calculates a fuzzy hash for each one having as a result a final hash combining them all.

\noindent{\textbf{Same language}.}
The next step is to determine if two non-equal pages are in the same language by comparing the language information we obtain in the data extraction step (see Section~\ref{sec:data-extraction}). 
If the language is different, we only compare the scheme structure of the page (since the content inherently differs).
When the language is the same, we rely on the output of the fuzzy hashing algorithm applied to the whole site (i.e., the entire HTML file). 
If we see that this HTML is the same, we determine that the two pages are mirrors. 
Instead, if the HTML files of the pages are different, we perform a further step.

\noindent{\textbf{Same language and different HTML}.}
We then apply a method to capture inconspicuous modifications.
These modifications may be attributed to changes in the products offered in the same market across the period of the crawl of the mirrors. 
Thus, we distinguish whether these differences stem from small changes in the scheme of the HTML (e.g., the CCS style), or in the content. 

As there are different types of reasons justifying inconspicuous modifications, we use a modular approach that can be fine-tuned to different scenarios. 
For this, we compare the pages using a weighted sum of the output of the fuzzy hashing algorithm for their {\em schemes} (only the HTML tags), and for their {\em content} (only the text). 
In particular, we apply two weights to the content and the scheme of a page (denoted as $W_c$ and $W_s$ respectively in Figure~\ref{fig:mirror}) that capture the relative importance each of these two elements has in the overall similarity measure.
The value for the weights and the threshold are established by empirical evaluation of the effectiveness of the algorithm to detect mirrors. 
Our implementation also empirically identifies that a similarity measure above 90\% implies that two pages are the same. 
When the similarity is below this threshold, we consider that the two pages have significant differences.

We detail empirical measures in Section~\ref{sec:analysis}, but we note here an important methodological step. 
We validate mirror detection using a custom tool that renders the HTML (without retrieving the media content, i.e., only the text). 
This lets us open the page in a browser and visualize potentially sensitive or even illegal content. 

\subsection{Network Analysis}\label{subsec:network_analysis}

To analyze the topology of Tor we use graph theory. 
For this,  we first create a graph that models the Page Linking relationship described in Section~\ref{sec:method:crawling}. 
In the graph, nodes represent singular pages and edges the relationship between pages (i.e., the pages are linked to each other). 
We next detail how we build and analyze the graph.

First, we build this graph ensuring that all links between the pages are added and linked using the same path as the crawling process. 
From the seeds to the endpoint pages, each site is connected based on the links referred between them. 

When a new page is crawled, the list of links is updated accordingly in the graph. 
Note that mirrors are not added to the graph to avoid redundant pages. 
However, we account for this information by annotating the page node with the mirrors it has. 
Several other features are added to each node of the graph to enrich the analysis:
\begin{inparaenum}
    \item title of the page,
    \item language,
    \item surface connection (boolean),
    \item depth (i.e., number of pages to reach a given one),
    \item timestamp,
    \item category,
    \item mirror existence (boolean),
    \item mirrors URL's.
\end{inparaenum}
Table~\ref{Table:graph_features} summarizes the features we use to annotate the nodes. 

Finally, through a reachability analysis, we compare the segment of the network we reach through the different seeds (e.g., OSINT or underground forum sources).
This lets us assess how effective the automatic seeding method is when compared to the manual one and how informative the different keywords extracted from underground forums are.

\begin{table}[h]
	\centering
    \caption{Features of a node in the Pages Graph.}
		\begin{tabular}{lp{6.5cm}}
 Feature & Description\\ 
\hline
\rowcolor[HTML]{EFEFEF} 
Title & The title of the page if it exists \\
Language & For analysis purposes related to statistics. Indicates \\
Surface & if the page is connected to the Surface Web or not \\
\rowcolor[HTML]{EFEFEF} 
Depth & Minimum path to reach this page from the seed \\
Timestamp & List of timestamps where the page was crawled/attempted \\
\rowcolor[HTML]{EFEFEF} 
Category & The type of content offered by the site \extended{as indicated by the classifier}\\
Has$\_$Mirror & Boolean value indicating whether a given site has mirrors \\
\rowcolor[HTML]{EFEFEF} 
Mirrors & List of mirrors that have not been included in the network \\ 
\hline
        \end{tabular}
	\label{Table:graph_features}

\end{table}

\subsection{Content Analysis}\label{subsec:methodContent}

One of the main goals of our study is to determine whether the type of content of the hidden services reached with Mimir is related to the seeds provided as input. This is of great interest to researchers (e.g., to understand the types of content being delivered in anonymous services) and to Law Enforcement investigators (e.g., to understand new forms of cybercrime or to identify sites trading deviant content).
To this end, we build a probabilistic model using Logistic Regression (LG). 
We choose LG due to its performance in other text-based learning tasks in comparison with other well-known algorithms that are good in this field as Naive Bayes, Random Forest, Decision Tree, or Support Vector Machines~\cite{pranckevivcius2017comparison}. \hl{To corroborate this, we explored the use of more complex algorithms (see Section} \S\ref{subsec:contentAnalysis}\hl{), obtaining similar accuracy at a higher latency.}
We use the categories in the Duta-10K dataset~\cite{duta10k-al2019torank} as ground-truth and build a multi-class classification algorithm. 

Duta-10K offers about 10K \texttt{.onion} sites labeled with 25 different categories that range from ``Counterfeit'' or ``Drugs'' to ``Market'' or ``Hacking''. 
The dataset was initially collected in 2017, and consequently, most of the 10K sites are no longer reachable. 
However, we got access to a large subset of raw HTML files from the authors of the dataset. 
We observe that some of the categories were \hl{underrepresented}, and thus we rely on the 11 (out of the 25) categories.
\hl{During training, we limited the number of samples to 200 per category to achieve a balanced dataset, using 200 samples when available or fewer when a category had less. 
This threshold was experimentally established to avoid biases from under-represented categories.
We excluded two types of categories: first, those with low samples count (e.g., Art, with 14 samples or Politics, with 2 samples), which were excluded due to insufficient support for consistent classification; and second, categories with a larger number of samples whose content overlaps with that of other categories, such as Personal (417 samples) and Services (284 samples).
Additionally, we removed the Empty category (1,350 samples), as it provided no useful content. 
Also, due to the nature of the ground truth, we limit the content analysis to only English sites. 
In total, we removed 2,314 samples (26.8}\%\hl{) and retained 6,321 samples (73.2}\%).

\begin{figure}[]
	\centering
	\includegraphics[width=\columnwidth]{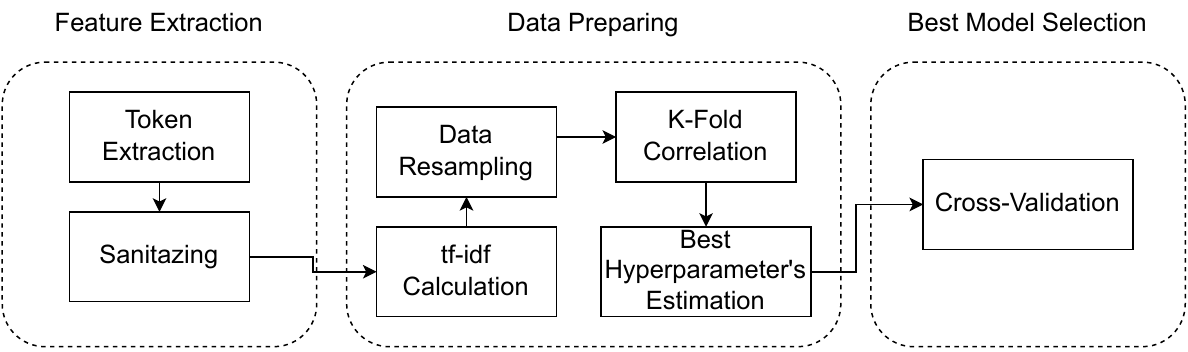}
	\caption{Content classification pipeline.}
	\label{fig:classifier_flow}
\end{figure}

The training process of the content classifier is divided into three tasks as shown in Figure \ref{fig:classifier_flow}: 
Feature Extraction, Data Preprocessing, and Best Model Selection.

\begin{enumerate}
\item {\bf Feature Extraction}. To extract features from the raw documents, the HTML tags are removed and the resulting content is turned into tokens. 
These tokens are then filtered by removing stop words, and the remaining tokens are lemmatized to convert each word to its inflected form or \textit{ ``lemma''}.

\item {\bf Data Preprocessing}. This process first calculates the Term Frequency-Inverse Document Frequency (tf-idf), which expresses the relevance of a word in a document from a set of documents, as defined by the following equation:

\begin{equation}
\texttt{tf-idf} = \frac{freq(term_{n}, document_{m})}{\log_2(\frac{\#documents}{\#documents\_with\_term_{n}+1)}}.
\label{tfidf}
\end{equation}

\noindent
The tf-idf is calculated using a Sublinear Term Frequency Scaling. This prevents repeated words from getting unnecessary relevancy, e.g., due to the existence of a list in the text or lack of vocabulary in the document. 
This means that the frequency is not natural but logarithmic. 
Then, we use bootstrapping to balance the training set. 
Finally, we shuffle the data and perform a division of the entire data set by using k-fold with k = 10.

\item {\bf Best Model Selection}. To estimate the optimal parameters of the Logistic Regression model,  
we perform a grid search over the parameters using a fraction of the training set and measure the penalty and the regulation strength: 
\begin{itemize}
    \item For the penalty, we take into account Lasso regularization (l1) which makes variable coefficients go to 0 in case they do not contribute as others do, ridge regression (l2) that decreases those coefficients but never lets them arrive at 0, elasticnet where l1 and l2 are added, and also we have check models without any penalty. 
    \item We also use the regulation strength (C) to determine the impact of the penalty on the coefficients, the stronger the regulation, the smaller the coefficients. 
\end{itemize}
\end{enumerate}

The three steps above are critical to find a proper trade-off between obtaining a general model or a fine-tuned model tailored to our training set. 
A poor regularization will make the model take into account non-relevant variables, whereas a too-strong regulation may weaken the model.

To assess the influence that the different parameters have over the performance of the model, we leverage the $F_{beta}$ score (with ${beta}$ equal to 1), defined by the following equation:
\begin{equation}
F_{\beta}=\frac{(1 + \beta^2) * Precision * Recall}{\beta^2 * Precision + Recall}.
\end{equation}

\noindent Our best setting uses a logistic regression with an {\em all vs all} strategy.

\subsection{Ethics}\label{sec:ethics}

{This research requires data collection from public sites. 
The researchers have experience dealing with the crawling of online underground communities hosted both on the surface and the Dark Web. 
We follow due precautions to minimize potential risks and harm to online users. 
First, the information we gather is obtained from public sources, which we reach by means of automatic navigation. 
The crawler is instrumented to gather textual content, and thus, we mitigate the risk of downloading illegal material such as indecent images of minors. 
Also, our analysis does not target specific groups or individuals and is conducted over aggregated data. 
Data is stored in an encrypted disk on our servers, and it is only accessible to the researchers involved in this project. 
Finally, following high ethical and legal standards, every evidence of illegal activities, especially those related to child pornography, is reported to the authorities. 
Overall, the benefits of our research outweigh the risks. 
We have obtained approval from our IRB and carefully considered Tor safety guidelines~\cite{safetyboard}.}

%% file: sections/analysis.tex
\section{Analysis}\label{sec:analysis}
\future{Correlate the money with sites categories}
In this section, we report our findings as a result of applying the methodology described above. 
First, we describe our experimental setup, including the description of the dataset collected. 
Second, we present the network analysis, and finally, the content analysis.

\subsection{Experimental Setup}
\label{subsec:experimentalSetup}
The dataset we use in this paper has been collected for a period of 8 months. 
We use a server with 24 cores and 47 GB of RAM. 
The crawler runs uninterruptedly except for some infrequent events (e.g., due to electricity cuts or server overloads).

The crawl starts after feeding the seeds to Mimir.
We collect nearly 7k seeds, out of which about 2\% stem from manual seeds (using OSINT searches), and the remaining are seeds collected automatically (using systematic queries to TSE using keywords from underground \hl{surface} forums). 
We see that 83\% of the \texttt{.onion} addresses in the initial set of seeds are unreachable. 
This highlights that the search engines and OSINT might provide outdated results, and motivates the use of online collection tools such as Mimir.
As a result, we discard unreachable \texttt{.onion} addresses and start the crawling process by visiting 1,157 seeds (17\% of the initial set).

Departing from the initial seeding, we reach 24.9k different sites. 
Table~\ref{table:dataset_char} offers a characterization of our datasets. 
We note also that the crawling workload grows almost exponentially as most of the content of the site includes a link to other sites. 

At the end of the 8-month crawl, the crawler has made a connection attempt to 34\% of the sites (5,002) in the \todolist{}. 
Likewise, we see that the crawler has gone through more attempts for a smaller proportion of URLs. 
Once the number of attempts reaches the maximum value set (i.e., 5), those URLs are appended to the \textit{Unreachable} list. 
In total, around 8K (21.4\%) of the URLs we attempt to connect to are currently unreachable, the large majority stemming from the set of living seeds. 

\begin{table}[h]
\centering
\caption{Dataset characterization. {\em Seeds} and {\em sites} (left to right) reached from the seeds.}
\begin{adjustbox}{width=0.4\textwidth,center}
\begin{tabular}{|cc|
>{\columncolor[HTML]{EFEFEF}}c ccc|}
\hline
\multicolumn{2}{|c|}{\cellcolor[HTML]{EFEFEF}\textbf{Seeds}} & \multicolumn{4}{c|}{\cellcolor[HTML]{EFEFEF}\textbf{Sites attempted/reached}} 
\\ 
\hline
\multicolumn{2}{|c|}{} & \multicolumn{1}{c|}{\cellcolor[HTML]{EFEFEF}} & \multicolumn{1}{c|}{} & \multicolumn{1}{c|}{\cellcolor[HTML]{EFEFEF}\textbf{Base}} & 4,121 \\
\cline{5-6}
\multicolumn{2}{|c|}{\multirow{-2}{*}{6,816}} & \multicolumn{1}{c|}{\multirow{-2}{*}{\cellcolor[HTML]{EFEFEF}\textbf{Accessible}}} & \multicolumn{1}{c|}{\multirow{-2}{*}{24,911}} & \multicolumn{1}{c|}{\cellcolor[HTML]{EFEFEF}\textbf{Mirrors}} & 20,790 \\ 
\cline{1-6}
\multicolumn{1}{|c|}{\cellcolor[HTML]{EFEFEF}\textbf{Accessible}} & \cellcolor[HTML]{EFEFEF}\textbf{Unreach.} & \multicolumn{1}{c|}{\cellcolor[HTML]{EFEFEF}\textbf{Unreachable}} & \multicolumn{3}{c|}{8,381 (sites)} \\ 
\cline{1-6}
\multicolumn{1}{|c|}{} &  & \multicolumn{1}{c|}{\cellcolor[HTML]{EFEFEF}\textbf{4 attempts}} & \multicolumn{3}{c|}{0} \\ 
\cline{3-6}
\multicolumn{1}{|c|}{} &  & \multicolumn{1}{c|}{\cellcolor[HTML]{EFEFEF}\textbf{3 attempts}} & \multicolumn{3}{c|}{5} \\ 
\cline{3-6}
\multicolumn{1}{|c|}{} &  & \multicolumn{1}{c|}{\cellcolor[HTML]{EFEFEF}\textbf{2 attempts}} & \multicolumn{3}{c|}{796} \\ 
\cline{3-6}
\multicolumn{1}{|c|}{} &  & \multicolumn{1}{c|}{\cellcolor[HTML]{EFEFEF}\textbf{1 attempts}} & \multicolumn{3}{c|}{5,002} \\ 
\cline{3-6}
\multicolumn{1}{|c|}{\multirow{-6}{*}{1,157}} & \multirow{-6}{*}{5,659 (seeds)} & \multicolumn{1}{c|}{\cellcolor[HTML]{EFEFEF}\textbf{Total}} & 
\multicolumn{3}{c|}{39,095} \\ \hline
\end{tabular}
\end{adjustbox}
\label{table:dataset_char}
\end{table}

\subsection{Network analysis}\label{subsec:an-network_analysis}

We first analyze the network topology. Concretely, we study the connectivity of the network, looking for Dark Web ``bubbles'', namely subgraphs --- i.e., a set of sites directly or indirectly connected to each other while being unconnected to the others. 

In particular, we see 1,040 subgraphs inside the network. 
We find that 99 of the subgraphs have 10 or fewer nodes, which shows that the Dark Web is highly fragmented and ``bubbles''  generally have a small size. 
Instead, we see that a small number of subgraphs have a large number of nodes that are all deeply connected. 
This shows that the topology of the Dark Web follows a power-law distribution much like the Surface on the advent of the Web~\cite{faloutsos1999power}. 
Table~\ref{table:topology} shows the top 5 subgraphs judging by the number of nodes, together with the number of edges.
We also show how many nodes have direct access to the surface web. 
All the subgraphs are weakly connected (i.e., not all pages link to all other pages). 

An interesting finding is that the largest subgraph represents $\approx$66\% of the \hl{total share of Tor that we reach}, being the Top 5 at 69\% of the network. 
Also, we observe that 9\% of the sites are directly connected from links on the surface. 
Our crawler only visits surface links when they are given as seeds.
This means that our systematic extraction of seeds lets us reach 1K subgraphs, with our crawler discovering most (91\%) of the Dark Web. 

\begin{table}[]
    \centering
    \caption{Size of the network and its top 3 largest subgraphs. \#LS is the number of nodes linked from the surface web.}
    \scalebox{1}{
    \begin{tabular}{l|rrr}
    & \textbf{\#Nodes} & \textbf{\#Edges} & \textbf{\#LS} \\
    \hline
    Full network   (1k subgraphs) & \multirow{1}{*}{4,121} & \multirow{1}{*}{3,360} & \multirow{1}{*}{412} \\
    \hline
    Subgraph 1 & 2,757 & 3,011 & 411 \\
    Subgraph 2 & 43 & 50 & 0 \\
    Subgraph 3 & 27 & 27 &  0 \\
    \end{tabular}
    }
    
    \label{table:topology}
\end{table}

Subgraphs of order 1 (i.e., just one node) represent the seeds that are unreachable. 
We can not confirm that these seeds will become available at a later stage, the same way that it is hard to get a static snapshot of the network due to the volatility of Tor. 
However, our results in terms of the coverage show that our methodology produces a more effective mapping of the Dark Web when compared to prior work (see Section~\ref{sec:related} for a detailed comparison with the related work). 

\subsection{Contribution of the seeding process}
\label{subsec:contributionSeeds}
We measure how our systematic seeding process improves the coverage of the network in terms of nodes found (i.e., onion sites). 
For this, we rely on the network analysis described above and 
perform an ablation study of the sources that seed our crawler. 
Recall that our seeding process stems from two sources: queries in Tor-focused search engines using keywords systematically extracted from underground \hl{surface} forums and manual entries obtained from OSINT.

We initially look at the \texttt{.onion} sites we reach from the manual seeding and take this set as a baseline. 
We then study the contribution that a keyword has to the coverage of the crawl over the baseline. 
For this, we first extract the subgraph of Tor sites that we visit because of all the manual seeds, namely Manual Seeds Subgraph ($MSS$). 
We then extract the subgraph that stems from each keyword seed $k$ and name it the Keyword Subgraph ($KS_k$). 
We finally compute the set difference (SD) as $SD = KS - MSS$, which retains only novel \textit{.onion} sites attributed to the keyword $k$, i.e., we remove all sites we reach through manual seeding (including the intersection). 

Table~\ref{table:keyword_contribution} shows the top 10 keywords that contribute the most to our crawl of the Dark Web. 
We observe that a single keyword like ``drugs'' or \hl{``hosting''} leads the crawler to reach as much as \hl{10.05\% and 9.29\% respectively} of the total gathered pages. 
This shows that crawling TSE with targeted keywords allows Mimir to obtain sizable improvements in terms of coverage when compared to the manual seeds. 
Table \ref{table:keyword_contribution} provides the contribution of each keyword individually with respect to the MSS. 
Since there is an overlap in the contributions of various keywords, the percentages in Table~\ref{table:keyword_contribution} are non-accumulative.

To understand the contribution of all the keywords as a whole, we repeat this process with $k = k_1	\cup k_2 \cup ...  k_n$, where $n$ is the total number of keywords. 
We denote this subgraph as All-Keywords Subgraph (AKS). 
\hl{As shown in the bottom row of Table} \ref{table:keyword_contribution}, the contribution to the coverage of all keywords together over the manual baseline is \hl{81.85}\%. 
In practice, this means that the number of \texttt{.onion} sites reached increases a \hl{177.55}\% when compared to manual seeding. 
Furthermore, the set difference between AKS and MSS is the empty set. 
This means that MSS is a subset of AKS, and thus, all the \texttt{.onion} sites we reach using the manual seeding method are covered by the automatic seeding approach. 
Plus, the automatic seeding approach covers a significantly larger portion of the Dark Web. 
All in all, \emph{this shows that our proposed automatic approach is an effective mechanism to avoid the cumbersome task of manually looking and entering a set of initial seeds or keywords.}
As all prior works require manual seeding, we see how our crawling strategy offers a competitive advantage in terms of scalability. 

\begin{table}[h]
\caption{\hl{Individual systematic seeding keyword nodes contribution per category.}}
\begin{adjustbox}{width=0.49\textwidth,center}
\begin{tabular}{c|cccccccccc|c}
 & drugs & free & hosting & software & hacking & forum & carding & counter & services & service & TCC \\ \hline
Counterfeit & 2.69\% & 1.70\% & 2.14\% & 1.70\% & 2.06\% & 2.35\% & 2.06\% & 2.26\% & 1.63\% & 1.33\% & 19.92\% \\
\rowcolor[HTML]{EFEFEF} 
Crypto & 1.75\% & 2.48\% & 1.55\% & 1.99\% & 1.46\% & 1.50\% & 1.63\% & 1.43\% & 0.51\% & 1.09\% & 15.38\% \\
Down & 0.73\% & 0.61\% & 0.63\% & 0.73\% & 0.44\% & 0.41\% & 0.58\% & 0.27\% & 0.46\% & 0.46\% & 5.31\% \\
\rowcolor[HTML]{EFEFEF} 
Drugs & 1.16\% & 0.53\% & 0.49\% & 0.58\% & 0.41\% & 0.41\% & 0.44\% & 0.39\% & 0.34\% & 0.39\% & 5.14\% \\
Forum & 0.29\% & 0.46\% & 0.41\% & 0.41\% & 0.36\% & 0.66\% & 0.05\% & 0.34\% & 0.32\% & 0.36\% & 3.66\% \\
\rowcolor[HTML]{EFEFEF} 
Hacking & 0.49\% & 0.53\% & 0.68\% & 0.68\% & 1.24\% & 0.56\% & 0.49\% & 0.46\% & 0.78\% & 0.68\% & 6.58\% \\
Hosting & 0.95\% & 1.29\% & 1.89\% & 1.12\% & 1.04\% & 0.99\% & 0.56\% & 0.41\% & 1.41\% & 1.12\% & 10.77\% \\
\rowcolor[HTML]{EFEFEF} 
Locked & 0.78\% & 0.56\% & 0.41\% & 0.49\% & 0.39\% & 0.53\% & 0.66\% & 0.29\% & 0.39\% & 0.41\% & 4.90\% \\
Market & 0.70\% & 0.49\% & 0.39\% & 0.39\% & 0.34\% & 0.41\% & 0.34\% & 0.46\% & 0.53\% & 0.39\% & 4.44\% \\
\rowcolor[HTML]{EFEFEF} 
Porn & 0.12\% & 0.78\% & 0.12\% & 0.10\% & 0.22\% & 0.12\% & 0.05\% & 0.39\% & 0.12\% & 0.22\% & 2.23\% \\
SN & 0.39\% & 0.44\% & 0.58\% & 0.53\% & 0.39\% & 0.27\% & 0.15\% & 0.24\% & 0.27\% & 0.24\% & 3.49\% \\ \hline
\rowcolor[HTML]{EFEFEF} 
AKS & 10.05\% & 9.85\% & 9.29\% & 8.71\% & 8.35\% & 8.23\% & 6.99\% & 6.94\% & 6.75\% & 6.70\% & 81.85\%
\end{tabular}
\end{adjustbox}
\label{table:keyword_contribution}
\end{table}

\hl{Similarly, we analyze how influential keywords contribute to the crawling of the different categories within the network. The rows in Table}~\ref{table:keyword_contribution} \hl{indicate the sub-graph of MSS considering only nodes of the given category (the last row being the full AKS graph). It can be observed that all the keywords primarily improve reachability to Counterfeit services, which is the most prevalent category in our dataset. After that, we observe interesting patterns since each individual keyword reaches the specific category related to it (e.g., drugs or hosting). 

While a keyword is not confined to influencing only its related category, it exerts the greatest impact within it. 
Note that multiple keywords can contribute to the same node.}

\subsection{Mirror analysis}\label{mirror_analysis}
As Section~\ref{subsec:an-network_analysis} analyzes the topology of singular pages in the Dark Web, we next evaluate our method to detect mirrors and provide a characterization.  
First, we perform an empirical analysis to set the scores $W_s$ and $W_c$ (introduced in Section~\ref{fig:mirror}) and we establish $W_s=0.3$ and $W_c=0.7$. 

We manually analyze English pages with at least one mirror (450 and 53.38\% in total) \hl{using a custom tool to render HTML sites without media content, thus preventing the exposition of the validator to indecent content}. 
During this step, we remove some (43) sites for several reasons, including wrong language detection and isolated problems rendering some of the HTML with the tool we develop for the validation.
All in all, we validate 408 (48\%) sites, of which only 22 are miss-classified, leading to a precision of \hl{97}\%.
Out of the 48\% sites, our algorithm detects 801 mirrors (355 identified with CTPH and 424 with MD5). 
Since hash match means that pages are exactly the same, we only need to manually evaluate the remaining 355.

Our crawler collects a total of 24,911 sites, out of which 83\% (20,790) are mirrors, 16\% (4,008) are unique \texttt{.onion} sites and  0.4\% (113) are  `unique' surface links wrongly provided as seeds at the beginning of the crawling process. \hl{Note that these unique sites are merely representatives of their respective mirrors. Thus, we define a unique site as the first intance of replicated site that we crawled, being the rest the mirrors.}
Note that the large majority of the links we see in our dataset are mirrors. 
Only 19\% of the unique sites have mirrors, and most of them (54\%) have only a single pair-wise match.  
There are also 11\% with 2 mirrors and 6\% with 3. 
The remaining 29\% have four mirrors or more. 
Interestingly, we spot 57 sites with more than 100 mirrors, with 2 sites having 1,126 and 1,203 mirrors respectively. 

More than half of the mirrors are exact copies (54.4\%), i.e., they have the same MD5 hash. The remainder (45.6\%) contains some minor modifications. Understanding the exact reasons for mirroring would require dedicated tools, and it is out of the scope of this paper. However, during our empirical evaluation of the mirror detection algorithm, we carried out manual validations over samples, which allow us to understand the differences. Most of the changes among the mirrors are minor and involve either the content, such as titles, or the HTML schema, such as updates in the design framework. Accordingly, we analyze two types of changes: one for easy-comparing content fields and another to analyze the modifications in the raw HTML.

 {We first run an automated analysis over sites detected as mirrors with CTPH} to count changes in the language and also identify information (i.e., cryptowallet addresses and email addresses). 
We find 15 pages in English that are also mirrored in other languages, mostly Italian, German, and French. 
There are 1,409 mirrors that have modifications in at least 1 Bitcoin address and one of them with at least one different Monero address. 
Regarding emails, there are 1,486 mirrors with at least one different email from one another. 
\hl{We hypothesize that minor changes may occur due to various reasons. First, some modifications may have occurred within the time lapse between the crawls of these mirrored sites. Second, it is possible that an unauthorized entity impersonated the legitimate site with malicious intent, potentially engaging in phishing or scams}~\cite{barr2020phishing}\hl{. And third, in cases where mirrors are legitimately managed by the same operator, there might not be an automatic synchronization process, causing some pages to remain outdated until the administrator updates them (if any).}
However, confirming these hypotheses is left for future work. 
 
\begin{figure}[h]
	\centering
	\includegraphics[width=.99\columnwidth, trim = 120 105 120 155, clip]{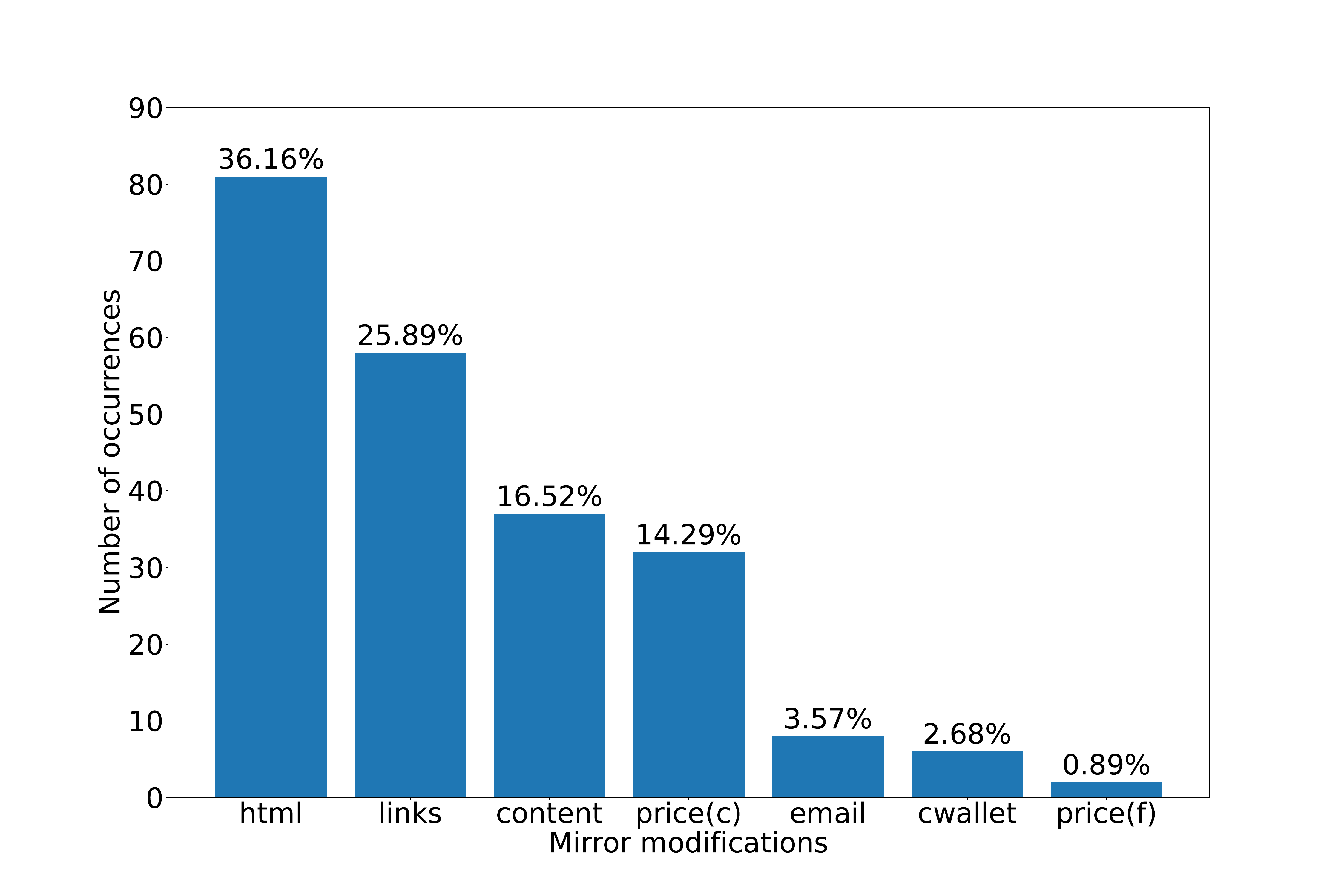}
	\caption{Minor modifications in the HTML of mirrored pages.}
	\label{fig:modifications}
\end{figure}

Second, we analyze differences in raw HTML files flagged as mirrors. 
Specifically, we manually analyze 216 pages in  English that have at least one mirror, together with their corresponding 335 CTPH-detected mirrors. 
The main modifications are seen in the HTML structure (content not included) as depicted in Figure \ref{fig:modifications}. 
The second most frequent difference relates to pages with changes in links of the site, followed by content modifications (only the text of the site). In some sites dedicated to trading, we see that the physical currency (e.g., dollar or euro) remains the same, but the equivalent price in currencies changes. 
We also see some pages with modifications in their cryptowallets. 
We see some sites generating one address per each purchase, and such changes should be studied carefully. 
Finally, we only see a few pages with modifications in the physical currency (FIAT money). 
Considering that pages are often crawled on different days, we see that some of the sites do not propagate changes to all mirrors automatically. 

\future{Put the font size of all figures larger, at least 24 points.}
The \texttt{.onion} addresses can be separated by versions, being their difference in length: v2 addresses are 16 characters, while v3 is 56 characters.\footnote{We note that, by the time our crawling started, v2 .onion addresses were not yet deprecated.}
In the 4,008 unique \texttt{.onion} sites, there are 1,061 which are v2 (26.4\%) and 2,795 v3 (69.7\%). There are also 152 URLs (3.8\%) that do not belong to either v2 or v3 and correspond to malformed URLs that, while resembling .onion addresses, link to a surface website. 
As for mirrors, there are 1,065 which are v2 (5.1\%) and 19,680 v3 (94.9\%), i.e., a larger proportion of the mirrors belong to the newest version.

Regarding connectivity, even though only 9\% (440) has direct access \textit{from} the surface (according to our initial seeds), 44\% of the sites reference a link \textit{to} the surface web.
The longest path from a seed to a site crawled, i.e., the maximum depth, is 6. However, this is just one exceptional case, and most of the pages are spread between the depths of 1 to 5. 
Specifically, there are 28\% (1,157) sites of depth 0 (i.e., seeds), 17\% (706) sites of depth 1, 8.2\% (340) sites of depth 2, 19.5\% (803) sites of depth 3, 21.8\% (902) sites of depth 4, (5.1\%) 212 of depth 5, and the one mentioned site in depth 6 ($<$1\%). 

\textbf{\hl{Mirror detection benchmark.}}
\hl{We build up a benchmark for comparing our algorithm with two of the most well-known state-of-the-art approaches, namely MinHash}~\cite{minhash} \hl{and SimHash. 
We first determine the optimal thresholds for each algorithm to conduct a fair benchmarking. 
To this end, we randomly choose 2,000 samples: 1,000 real unique-mirror pairs and 1,000 pairs of pages that are not mirrors.
We then calculate the number of correctly detected mirrors and non-mirrors for both methods across 10 different thresholds, ranging from 0 to 1 in increments of 0.1. This experiment indicates the optimal threshold score for each method, i.e., 0.8 for SimHash, and 0.4 for MinHash.}

\begin{table}[h]
\caption{\hl{Duplication Algorithms Benchmark}}
\begin{adjustbox}{width=0.5\textwidth,center}
\begin{tabular}{|c|c|c|c|c|c|c|c|}
\hline
\rowcolor[HTML]{EFEFEF} 
\textbf{Method} & \textbf{TP} & \textbf{FP} & \textbf{FN} & \textbf{Precision} & \textbf{Recall} & \textbf{F1-Score} & \textbf{Repetitions} \\ \hline
Mimir           & 10,498       & 9           & 3           & 0.99               & 0.99            & 0.99              & 0                    \\ \hline
SimHash         & 10,495       & 15,440       & 6           & 0.40               & 0.99            & 0.57              & 15,336                \\ \hline
MinHash         & 10,499       & 6,634        & 2           & 0.61               & 0.99            & 0.75              & 6,579                 \\ \hline
\end{tabular}
\end{adjustbox}
\label{table:hash_benchmark}
\end{table}

\hl{Table} \ref{table:hash_benchmark} \hl{summarizes the results of the benchmark. The three methods obtain similar True Positive (TP) and Recall rates. 
However, a key issue arises with the False Positive (FP) rate. SimHash and MinHash frequently identify the same site as a mirror for multiple representative pages, significantly reducing their precision. 
In contrast, Mimir stands out for its robustness in avoiding this issue, demonstrating greater resistance to incorrectly matching the same page as a mirror for different representatives. 

Overall Mimir is particularly effective for measuring HTML document similarity. 
While SimHash and MinHash are good approaches, they have limitations with non-nearly identical pages. 
SimHash uses hamming distance for detecting near-duplicate detection small changes that may affect the similarity score. 
MinHash's reliance on text tokenization may introduce excessive noise due to HTML redundancy, reducing its effectiveness when using Jaccard similarity as a metric.}

\subsection{Content analysis}\label{subsec:contentAnalysis}

One of our key goals is to analyze the content hosted in the Tor network in order to localize cybercrime-related HS. For such purpose, we first look at the language used. 
Then, we categorize the content, and finally, we detect the mirrors.
Overall, there are 44 languages. 
We see that 88\% of the pages are written in English. The remaining 12\% are widespread in terms of language since each one accounts for less than 1\%. These results are consistent with previous studies on the Tor network~\cite{avarikioti2018structure}, and confirm that most of the content is still offered in English. 

\alfonso{\hl{ALL THIS PART IS NEW}}

\textbf{\hl{Model Benchmark:}}
\hl{For comparison, we trained various machine learning models using the same training set. Apart from Logistic Regression (LR), we select state-of-the-art algorithms for text classification, i.e., Convolutional Neural Networks (CNN), Bidirectional Encoder Representations from Transformers (BERT), and a recent variant focused on Dark Web content, DarkBERT}~\cite{darkbert}.

\begin{table}[h]
\caption{\hl{Duplication Algorithms Benchmark}}
\begin{adjustbox}{width=0.5\textwidth,center}
\begin{tabular}{|c|c|c|c|c|c|c|}
\hline
\rowcolor[HTML]{EFEFEF} 
\textbf{Model} & \textbf{Accuracy} & \textbf{Precision} & \textbf{Recall} & \cellcolor[HTML]{EFEFEF}\textbf{F1} & \cellcolor[HTML]{EFEFEF}\textbf{F2} & \cellcolor[HTML]{EFEFEF}\textbf{Time} \\ \hline
LG & 0.863 & 0.757 & 0.864 & 0.798 & 0.833 & 0.0169 \\ \hline
CNN & 0.834 & 0.699 & 0.690 & 0.691 & 0.689 & 1.2637 \\ \hline
BERT & 0.847 & 0.682 & 0.697 & 0.685 & 0.690 & 15.3399 \\ \hline
DarkBERT & 0.862 & 0.704 & 0.712 & 0.704 & 0.708 & 46.0361 \\ \hline
\end{tabular}
\end{adjustbox}
\label{table:ml_benchmark}
\end{table}

\hl{Table }\ref{table:ml_benchmark} \hl{shows the results of our benchmark experiment. 
The CNN, BERT, and DarkBERT achieve similar accuracy to that of LG. 
However, the three models have suboptimal precision and recall. 
Additionally, the time required to classify the test set increases significantly in comparison with Logistic Regression. 
Simple models like Logistic Regression and Naïve Bayes often outperform complex models like BERT or CNNs in text classification when data is limited. 
These traditional models, relying on hand-crafted features like TF-IDF and word n-grams, excel in specific tasks without needing long-term dependency understanding}~\cite{illegal_darknet, power_simplicity, churn}. 
\hl{In contrast, deep models tend to overfit without large datasets, learning irrelevant patterns or missing the correct ones.
Overall, for our specific task and dataset size, Logistic Regression outperforms the rest of the models.}

\vspace{.15cm}
\noindent\textbf{Categories}.
We built a custom ML classifier (see Section~\ref{subsec:methodContent}). 
The training has been performed with the set of labeled pages from the Duta-10K dataset~\cite{duta10k-al2019torank}. From the total list of 10,367 URLs, only 1,099 (10\%) were still available, leading to an imbalanced dataset. 
After balancing the data and removing under-represented categories, the final training set accounts for 2,200 sites and 11 categories (200 samples per category).
The final test set contains 5,628 pages, offering an accuracy of 86\% (75\% precision and 86\% recall). 
Overall, we get 79\% F1 and 83\% F2 scores.

The trained model is used to classify all sites. 
In addition to that, we also measure the reliability of the classifier by using conformal evaluation to calculate the percentage of the unreliability of a given classification, i.e., the difference between the most probable class and the second one, which can be expressed as: 
$1-(P(C_{1^{st}})-P(C_{2^{nd}}))$.

According to this metric, we find that the reliability of more than half (52.52\%) of the pages was higher than 90\%, with $\approx$21\% being reliable at 99\%. Only 8.6\% of the pages were below the 10\% of reliability, followed by a $\approx$12\% that are between 11\% and 20\%. Therefore, it shows the classifier is not only efficient but also reliable. 

\begin{figure}[h]
	\centering
	\includegraphics[width=.99\columnwidth, trim=36 0 38 10, clip]{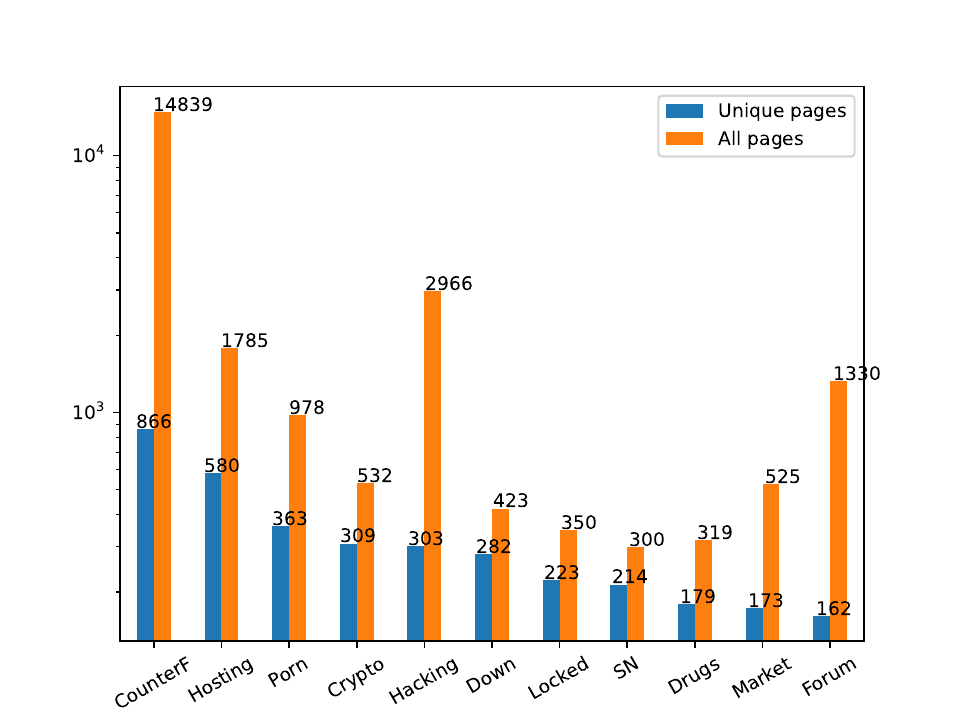}
	\caption{Histogram representing the distribution of sites per categories with and without filtering out mirrors. 
	}
	\label{fig:categories_dis}
\end{figure}

Using the trained model, we infer the category for all the English sites in our dataset. 
Figure~\ref{fig:categories_dis} shows the resulting categories. 
For a better understanding, these are described in Table \ref{Table:categories_description}. We indicate which categories might be related to \hl{deviant} activities (e.g., Drugs or Locked). We note, however, that some of the others can also contain cybercrime-related content, e.g., crypto, market, forum, or porn (as we show in \S\ref{sec:use_case}).
We see that the predominant class is ``Counterfeit'', which mainly aims to sell stolen credit cards. 
It is followed by the ``Hosting'' category which is related to Servers, File Sharing, and Links Directories.
The third category with more matches is ``Porn'' which provides explicit sexual content.
In total, we see 1,571 (38.12\%) unique pages whose categories inherently define \hl{deviant} content (i.e., Counterfeit, Drugs, Hacking, and Locked).
However, we see 18,474 (74.4\%) if we take into account the mirrors. 
Table~\ref{Table:cybercrime} offers a breakdown of the prevalence of cybercrime in our dataset per category.
{We note that these figures only offer a lower bound as other HS may have \hl{deviant} content also (e.g., ``Porn'' or ``Markets'')}.

\begin{table}[h]
\centering
\caption{Detailed cybercrime related sites numbers.}
\scalebox{.72}{
\begin{tabular}{c|c|c|c|c|c|}
\cline{2-6}
                                                               & \cellcolor[HTML]{EFEFEF}\textbf{Counterfeit} & \cellcolor[HTML]{EFEFEF}\textbf{Drugs} & \cellcolor[HTML]{EFEFEF}\textbf{Hacking} & \cellcolor[HTML]{EFEFEF}\textbf{Locked} & \cellcolor[HTML]{EFEFEF}\textbf{TOTAL} \\ \hline
\multicolumn{1}{|c|}{\cellcolor[HTML]{EFEFEF}\textbf{Unique}}  & 866 (55.2\%)                                & 179 (11.4\%)                          & 303 (19.3\%)                            & 223 (14.2\%)                           & 1,571 (38.1\%)                         \\ \hline
\multicolumn{1}{|c|}{\cellcolor[HTML]{EFEFEF}\textbf{Mirrors}} & 13,972 (82.9\%)                              & 132 (0.8\%)                           & 2,652 (15.7\%)                           & 84 (0.5\%)                             & 16,840 (81.3\%)                        \\ \hline
\multicolumn{1}{|c|}{\cellcolor[HTML]{EFEFEF}\textbf{TOTAL}}   & 14,839 (80.3\%)                              & 319 (1.7\%)                           & 2,966 (16.1\%)                           & 350 (1.9\%)                            & 18,474 (74.4\%)                        \\ \hline
\end{tabular}
}

\label{Table:cybercrime}
\end{table}

\begin{table}[h]
\centering
\caption{Classifier categories description, with (*) denoting cybercrime-related activities.}
\scalebox{.85}{
\begin{tabular}{|>{\columncolor[HTML]{EFEFEF}}c|p{7cm}|}
\hline
\textbf{CATEGORY} & \multicolumn{1}{c|}{\cellcolor[HTML]{EFEFEF}\textbf{DESCRIPTION}} \\ \hline
\textbf{\begin{tabular}[c]{@{}c@{}}Counterfeit*\end{tabular}} & \begin{tabular}[c]{@{}l@{}}Carding (i.e. credit card), Fakes ID and Money sales.\end{tabular} \\ \hline
\textbf{Crypto} & Cryptocurrency buying, selling and  mining \\ \hline
\textbf{Drugs*} & Drug marketplaces or other related sites \\ \hline
\textbf{Forum} & Discussion forums of any type. \\ \hline
\textbf{Hacking*} & Hacking services or pages related to hacking. \\ \hline
\textbf{Locked*} & Log in, sites closed by the authorities and so on. \\ \hline
\textbf{Down} & Pages that are not available anymore. \\ \hline
\textbf{Market} & Different marketplaces as virtual items, weapons, pharmacy... \\ \hline
\textbf{Porn} & Pages related to pornography. \\ \hline
\textbf{\begin{tabular}[c]{@{}c@{}}Soc.-Network (SN)\end{tabular}} & Blog, Chat, Email. \\ \hline
\textbf{Hosting} & File-Sharing, Folders, Search-Engine, Server, Directory. \\ \hline
\end{tabular}
}
\label{Table:categories_description}
\end{table}

\noindent\textbf{Mirrors}.
Figure~\ref{fig:categories_dis} shows with the orange bar the number of pages (including mirrors) per category. 
If we were not to filter mirrors out, we may observe some important differences. First, some classes are really over-represented as ``Counterfeit'' which the number of pages appears to be 1,713\% higher if we do not identify which of those sites are mirrors. Second, if we want to understand which content represents the cybercrime in the Tor Dark Net better at some point in time, this over-representation may lead to misunderstanding the real picture of the network. For example, the category ``Hacking'' may look like it is the second service more extended in the network, but in reality, it holds the 5$^{th}$ position, and which is more interesting as well as ``Forum'' which seems to be fourth when actually is the last one.
This shows that accounting for mirrors when analyzing the Dark Web is important to avoid biases in the interpretation of the results, and it can also help law enforcement to make diligences more effective. For example, in the case of Counterfeit, only 5.8\% of the sites require in-depth inspection (i.e., the unique ones), 10\% in the case of Hacking sites, or 56.11\% in terms of Drugs.
n
Our results also open new research questions on the type of pages that are being mirrored, and for what purpose. 
A deeper analysis looking at the raw HTML of the most mirrored Top-10 unique pages shows that 70\% of them were ``Counterfeit''. 

%% file: sections/use_case.tex
\section{Use case: distribution of child abuse material}\label{sec:use_case}

A staggering type of abusive content that finds shelter in the Dark Web is the distribution of child abuse material. 
We next see how our contribution has been helpful in the process of finding child abuse-related \texttt{.onion} sites, all of which have been reported to our national Law Enforcement contact.

During the manual validation in Section~\ref{mirror_analysis}, we identified 20 sites hosting child abuse content \hl{from the samples  categorized as Porn using our ML classifier}. 
We note that our manual inspection is restricted to text only. 
Our crawler does not download media content, and thus our manual validation went only through the text we extract offline from the HTML files we collected, and never through images or videos.
The size of this set (20 sites) is too small to include child abuse as a category of the ML classifier used for content analysis (c.f. Section~\ref{subsec:contentAnalysis}). Thus, we developed a custom keyword search algorithm for this. 
Our algorithm flags \texttt{.onion} sites that contain child abuse-related keywords. 
We extract these keywords from the 20 manually detected child abuse sites that capture the jargon used by pedophiles and that it is exclusive to child abuse. For ethical reasons, we do not disclose such keywords since these would facilitate miscreants to reach these pages using a keyword search with Tor search engines.

Many forums and collaborative sites on the Dark Web explicitly prohibit child abuse content.
In general, site administrators and moderators that forbid this type of content use short and clear statements. 
We identify these cases using formal language modeling and use it to avoid false positives.
In particular, we model every sentence and extract the structures for adjectives and verbs. 
We then model negative words around our keywords or explicit forbids following the structure we define in Table~\ref{tab:structures}.

We use four general rules to model adjectives and verbs around the keywords of interest. 
Sentences can be positive or negative. 
In the case of verbs, negative means that they express prohibition, and positive verbs express the opposite. 
As prohibition can be manifested as positive and negative sentences, we model both cases and account for the meaning given by the adjective/verb as well as for negative abbreviated forms like \textit{``n't''}. 

\begin{table}[h]
\centering
\caption{Structures to detect child abuse content allowance. KEY: Keywords, VBZ: Infinitive verb, VBN: Past participle verb, JJ: Adjective, WILL, NOT/N’T: Specific words. The symbols "+" and "-" denote positive and negative words, with verbs indicating allowance or prohibition.}
\begin{adjustbox}{width=0.4\textwidth,center}
\begin{tabular}{|c|c|}
\hline
\rowcolor[HTML]{EFEFEF} 
\textbf{Structure}       & \textbf{Example}              \\ \hline
KEY + VBZ + NOT/N'T + (+)VBN & {[}keyword{]} is not allowed  \\ \hline
KEY + WILL + VB + (-)VBN & {[}keyword{]} will be removed \\ \hline
KEY + VBZ + NOT/N'T + (+)JJ  & {[}keyword{]} is not welcome  \\ \hline
KEY + VBZ + (-)VBN       & {[}keyword{]} is censored     \\ \hline
KEY + VBZ + (-)JJ        & {[}keyword{]} is forbidden    \\ \hline
\end{tabular}
\end{adjustbox}
\label{tab:structures}
\end{table}

We apply the algorithm to the sites that have been classified in the category `Porn', and find 180 pages potentially distributing child abuse material. 
We perform a validation of all 180 hits (again, only judging the text), and confirm that \hl{165 unique sites (91,66\% accuracy)} claim to allow the distribution of indecent images or videos with minors.
We attribute the errors to some expressions that were not modeled and to the lack of some keywords in the child pornography lingo.
Notably, after the validation process, we examined additional mirror sites identified by Mimir, increasing the detection count to 505 pages (a 306\% increase).

Whereas we rely on a dataset collected up until April'22 (c.f., Section~\ref{subsec:experimentalSetup}), we observe that 78 of these sites (49\%) were operative in December'23, \hl{and that 45 are still operative by March'25} --- at the time of this writing.
Alarmingly, we found a hidden service (which was mirrored in 4 .onion sites) that provides interactive pornographic content with minors through live cams which, if confirmed, would pose systematic aggression and possible child trafficking. Unfortunately, this site is still online in 2 out of the 4 mirrored sites in \hl{March'25}, almost three years after we first observed it. \hl{This suggest that, even if Law Enforcement is able to take down some of these sites, the service remain operative as it is mirrored in other sites. This highlights the importance of considering mirrors when analysing the Dark Web.}
\hl{Accordingly, }all these sites have been reported to our national Law Enforcement contact point and are currently under investigation. Preliminary feedback confirmed that many of them actually offer illegal content.

%% file: sections/discussionLimitations.tex
\section{Discussions}
\label{discussions}

This work provides new insights into the Tor Network. In this section, we discuss the main limitations of our work together with key takeaways. 

\subsection{Limitations}
Measuring the topology and contents of Tor is challenging and requires dedicated tools that have limitations. 

\noindent\textbf{Coverage.}
Our analysis is bounded by the coverage of our crawler. 
This is a limitation prevalent in previous work, although our crawling methodology is more exhaustive. 
In particular, we crawl 233\% more reachable sites than M. Bernaschi et al.~\cite{bernaschi2022onion} for as twice as long (see \S\ref{sec:related}).
\extended{Despite efforts to maximize the coverage of the crawling process, we emphasize that Tor is an extremely volatile network. 
Our work attempts to connect to every unreachable time several times at different points in time. 
However, we report results through the lens of a static dataset while our crawling was still active. 
This means that we might have different network topologies (e.g., different number of `bubbles') after we crawl the continuum of the URLs in the \todolist{}. 
We plan to address this as part of our future work by devising probabilistic estimators of change based on the interpolation of measures taken after monitoring the volatility of Tor over time.}

The novel systematic extraction of seeds offers a key advantage with respect to prior work in terms of scalability, which translate into larger coverage without the need for manual seeds crafted by experts.

We note that our dataset represents a static snapshot of the network (i.e., the one we got in our crawling period). 
Having dynamic information requires living systems to continuously conduct crawls, a capability that the proposed crawler Mimir contains.
However, we opted to stop the crawling after one snapshot was retrieved following Ethical advice. 
Our goal is to provide a singular content and network characterization of Tor, and not study its temporal evolution.
Thus, our research effort focuses on performing a single crawl and does not require a production system that performs a continuous crawl. 
This way we avoid increasing the overall workload on the Tor network.

\noindent\textbf{Ground-truth.}

Another important limitation stems from the lack of ground truth to build a content classifier. 

Our work relies on the annotations made as part of the Duta project \cite{duta6k-al2017classifying,duta10k-al2019torank}.

However, some categories were \hl{underrepresented} in the data set and we thus remove them. 
We showed that for the categories we retain we are able to obtain reliable performance. 
The lack of good quality ground truth is a prevalent problem in cybersecurity. 

\hl{As part of our future work, we aim to expand the number of categories used in content characterization. 
However, manually labeling content is both time-consuming and raises important ethical and legal concerns.
It is important to note that content analysis is not a primary contribution of this paper. 
Rather, it is used to validate the effectiveness of Mimir in discovering pages related to the provided keywords. 
In fact, since Mimir is a modular architecture, integrating other machine learning models is straightforward.}

\noindent\textbf{Rich- and media-content.}
Our work primarily uses text to identify mirrors and classify the content of the sites. 
Ethical and legal concerns have prevented us from downloading multimedia content from Dark Web sites that may contain deviant content (e.g., child abuse material, actionable exploits for de-anonymization\cite{conti2016selfrando}, or malware \cite{6838909}).
We leave for future work the use of ethical and legal methodologies to analyze this content without harming researchers, e.g., automatic image processing. 
Different from the clearnet~\cite{pastrana2019measuring}, our study entails a higher risk of downloading illegal content, which brings unique challenges that require careful consideration and active cooperation with law enforcement agencies for this task.

\noindent \textbf{Landing page scope.} While vertical crawling may offer more in-depth insights, since our main goal is to discover as many hidden services as possible, we set a horizontal crawling approach, not only to limit the cost in time but also the bandwidth exposed to the Tor network, which is quite important when using the Tor protocol. 
Our results in category classification and mirror assessment have shown that data beyond landing pages is not essential in this task except for cases where the landing page is a login page or has a Captcha due to the limited textual content available for classification and the structural similarity of the data.
Such cases require tailored efforts, which are out of our scope.
Furthermore, our ethics protocol steems us to sign in to sites that may offer child abuse content, which limits our ability to further explore this conclusion.
We see that there are not many such cases, two (0.48\%) captchas and four (0.97\%) cases with a Login form.
Vertical crawling would introduce an important overhead to the Tor network, jeopardizing the objectives of this work.

\noindent\textbf{Mirror detection algorithm and copycat sites.}
We correctly identify the 97\% of Mirrors. 
The 3\% misclassification is primarily due to two reasons: small HTML files with minimal content and sites that use the same framework but in different configurations (e.g., a personal email server). However, differentiating between original sources and copycat mirrors is challenging, especially given Tor's anonymity features. 
Minor differences between mirrors are insufficient for attribution. 
Reliable attribution requires a deeper analysis beyond the scope of our contribution.

\subsection{Key Findings}

While our study presents some limitations as discussed above, we offer a unique analysis of singular services on the Dark Web, \hl{with a special focus on the impact of duplicate content in the network.}

Our findings constitute a large longitudinal crawl that requires a comparatively small number of seeds. 
We next summarize the key takeaways of our work and discuss their implications for research.

\noindent\textbf{By leveraging an innovative method for seeding and crawling, we were able to reach a larger proportion of the Tor Network.}
Using trending topics in underground surface forums is a promising direction \hl{when seeking deviant content in Tor}. 
This method reaches over twice the number of onion sites compared to manual seeding without missing any. \extended{Future work will explore adding a feedback loop from .onion sites to the seeding process.}
Furthermore, by making our crawler aware of the network context, we reduce the impact of Tor's volatility, leading to further increased page reach. 
As a result, Mimir significantly reduces the time required to reach hidden services. 
Additionally, our horizontal crawling strategy shows better network coverage with a higher number of sites per URL visited compared to prior work. 
In particular, Mimir has discovered 25k pages from 1k seeds, achieving an amplification factor of 2.2.

\noindent\textbf{\hl{Our analysis confirms that Tor needs to be studied as a dynamic network, due to the high volatility of its content}}. 
\hl{Depending on the crawling period (and even the time when a page is visited), the connectivity patterns change. 
Our crawler makes an important contribution to the study of Tor. 
We can not claim robustness on our network analyses since we ignore whether the connections observed during our crawling period will remain available in the near future. 
However, our crawling methodology is an important stepping stone toward understanding the dynamics of the Dark Web, exchanging the limitation that this entails for an advantage}.

\noindent\textbf{Mirror matter when analyzing \texttt{.onion} sites content.} 
{A remarkable takeaway from our study, achieved by the proposed detection algorithm, is the amount of replicated content (mirrors) present in Tor.}
Around 82\% of the landing pages accessed through our crawler are replicated content.  
We emphasize that this figure includes exact copies, and also pages with minor modifications in the landing pages.
Our results comparing \hl{Hacking against Porn or Cryptocurrencies} pages show the limitations after not considering mirrors. 
Without mirror detection, \hl{hacking-related} pages appear as the most prevalent category (inflating their importance over personal-related sites by $\approx$2,966\%). 
When looking at singular pages instead, \hl{Porn} pages are a more prevalent category \hl{followed by Cryptocurrencies}. 

Thus, confirming that factoring in mirrors is essential, which was a common gap in the existing literature. 
Accounting for mirrors can also help make informed decisions when prioritizing actions designed to thwart these activities, making prosecution more effective.
 
\noindent\textbf{Our experimental work lets Mimir focus the search on cybercrime.}
Our analysis reveals that more than 74\% of the sites in our dataset relate to cybercrime or \hl{deviant} activities. 
The remaining sites may or may not contain such activities, but by itself, it demonstrates that Mimir focuses on cybercrime.
There is a significant presence of ``Counterfeit'' and ``\hl{Porn}'' sites. 
The former provides trading material for various illegal activities such as carding, fake IDs, and money counterfeiting, while the latter offers \hl{both licit and illicit pornographic content.}
\hl{Together with counterfeit, hacking services stand out to be the most replicated HS for victimizing individuals through means such as credential theft or Denial of Service (DoS) attacks. }
It is worth noting that if we consider only unique sites, the percentage of cybercrime-related sites decreases to approximately 38\%. 
However, the fact that 81.32\% of the mirrors are related to cybercrime highlights the substantial effort that cybercriminals put into expanding their reach and attracting more clients. 
This suggests that the primary purpose of mirrors is to increase their visibility in the network, and hence, gain more market share.

Moreover, we conducted a use case focused on detecting child abuse material, achieving an accuracy of 88.33\% in their detection. 
Specifically, we departed from an especially scarce dataset of just 20 manually detected child abuse sites and we devised a method to reliably identify child abuse material. 
We were able to detect \hl{159 unique sites (and 505 mirrors)} pages containing such material with an accuracy of \hl{91.66\%}, out of which  45 (28.3\%) were still operative from the same \texttt{.onion} address at the time of this writing. These sites have been reported to Law Enforcement and are currently being investigated.

\textbf{\hl{Applications:}}
A key contribution of our study is that we show that guiding a Tor crawler from the beginning, using a set of keywords, amplifies its efficiency to reach sites from areas of interest. Mimir enhances the search on specific topics or threats and enables research in the desired specific domain. This approach helps to prioritize the discovery of relevant sites, ensuring that critical targets can be identified quickly and comprehensively. 
Besides the mentioned benefits for Law Enforcement when prosecuting illegal activities, Mimir can assist in informing appropriate cybersecurity policies, e.g., facilitating studies on current cybersecurity threats, vulnerabilities, and attack trends. In this regard, mirror detection plays a crucial role in enhancing threat intelligence and stakeholder analysis.  By identifying multiple mirrors of a site, investigators can ensure continuous monitoring of deviant activities, even if some of the domains are taken down or moved. 

While mirror sites amplify the detection coverage, they can create challenges when interpreting data. Indeed, we believe that our study should drive researchers to get more accurate insights by considering the noise and nuances of mirrors in Tor measurements.
Without considering mirror effects, analysis may result in a distorted view of the dark web ecosystem. 
This, in turn, may lead to incorrect conclusions (e.g., an increase in a specific type of malware or a surge in demand for a particular illegal item), causing necessary actions to be overlooked due to decision fatigue or cognitive overload.

%% file: sections/relatedWork.tex
\section{Related work}\label{sec:related}
Previous work on Tor has focused on understanding criminal activities, though often looking at particular sites, e.g., forums or markets~\cite{soska2015measuring,nazah2020evolution,van2018plug,labrador2022examining}. 
Other works have also studied the entire network ecosystem, focusing on particular activities and looking at the topology of Tor and its content. 
We next provide an overview of the most relevant ones, which we summarize in Table~\ref{tab:related}. 

\begin{table}[h]
    \centering
    \caption{Related work compared for crawling period, \#seeds, crawl depth, URLs seen, Hidden URLs (\# HS), Hidden Second-Level Domains (\# HSLD) analyzed, reachable (200), and categories analyzed (\# Cat). Symbol – indicates unavailable or unclear information.}
    \scalebox{.75}{
    \begin{tabular}{|c|ll|c|c|c|c|c|c|c|}
    \hline 
         & \multicolumn{2}{c|}{\bf Period (until)} & \bf \#Seeds & \bf Depth  & \bf \#URL & \bf \#HS & \bf \#HSLD & \bf 200  & \#\bf{}Cat  \\
    \hline
    \hline
      
      \cite{faizan2019exploring}   & \multicolumn{2}{l|}{1 month (m)} & --  & 1 & --  & -- & 26K & 2,125 & 31 \\ 
    \hline
      \cite{avarikioti2018structure}   & \multicolumn{2}{l|}{1 week} & 20K & max & 7-67M &  \multicolumn{2}{c|}{34K} & 7,566  & 30 \\ %
    \hline
    
    \cite{duta6k-al2017classifying} & 2m & (07/'16) 
    & -- & 2 & -- & \multicolumn{2}{c|}{250K}  &  7,931  &  26 \\
    \hline
    
      \cite{duta10k-al2019torank}  & 2m & (07/'17) 
      & -- & 2 &  -- & \multicolumn{2}{c|}{\cite{duta6k-al2017classifying} + 125K}  &  10,367 & 25 \\    \hline
      \cite{bernaschi2019spiders}  &  \multirow{2}{*}{4m} &  \multirow{2}{*}{(05/'17)} & 
    
      \multirow{2}{*}{--} & \multirow{2}{*}{--} & \multirow{2}{*}{--} &  \multirow{2}{*}{3M} & \multirow{2}{*}{30K} & \multirow{2}{*}{10,685} & -- \\
       \cite{bernaschi2022onion}  & & &  &  &  &   &  &   & \cite{duta6k-al2017classifying} \\ 
       \cite{zabihimayvan2019broad} & 1m & (07/'18) 
       & 20K & 4  &  1.2M & 40K  & 1,766 & 7,782 & 9 \\
    \hline
    \cite{Burda19} &
    4m & 
    (01/'19) &
    - &
    - &
    144.5k &
    20.4k &
    - &
    7,831 &
    6\\
    \hline
    \hline
      \bf Us  & 8m & (05/'22) 
      & 6,816 & 1 & n/a & n/a & 50,294 & 24,911 & 12 \\
    \hline

    \end{tabular}
    }
    \label{tab:related}
    \vspace{-1.5em}
\end{table}

\noindent{\bf Crawling.}
Al-Nabki et al. crawled in 2017 over 250k sites\extended{ in a period of 2 months}~\cite{duta6k-al2017classifying}. 
Then, they manually labeled sites into \hl{24} classes, leading to the DUTA dataset with 7k sites available, which we use in our study. 
DUTA was later updated in 2019~\cite{duta10k-al2019torank} --- leading to 3k new addresses\extended{, which were labeled and added to the previous dataset} --- and used recently in 2022~\cite{bernaschi2019spiders,bernaschi2022onion}. 
However, these works do not account for mirrors, as evidenced in our paper, which are a prevalent phenomenon in Tor.

\extended{\noindent{\bf Tor network.}
Later, \cite{avarikioti2018structure} analyzed the topology of Tor and the content of 10k sites. 
They manually labeled a subset of 200 sites and applied Active Learning for the content classification using SVM. 

Their seeds stem from several sources, reaching around 20K sites.
 
They also performed a language and content analysis with 30 categories separating their activities as legal and illegal pointing out that at this time, most of the content was labeled as ``Empty''.
\hl{This means that most pages do not have enough content to be classified.}

Recently, \cite{faizan2019exploring} studied 25k URLs (6k available). 
They classified sites into 31 categories and differentiated them into legal and illegal. 
While they note the presence of clones, mostly related to Bitcoin and forged documents, they neither specify how many nor how they found them. 

In 2022, authors in~\cite{bernaschi2019spiders,bernaschi2022onion} studied the Tor network using graph analysis. 
To this end, they built a crawler that was used to collect nearly 30k hidden services and conducted three crawls, showing the high volatility of the services. They also used the DUTA dataset to build a classifier.
However, they have not accounted for mirrors, and thus, the results might be biased.}

\noindent{\bf Mirrors.} 
{Our work is motivated by prior work observing mirrors in Tor}~\cite{Burda19,wang2023comprehensive} { in just 7.8k} \texttt{.onion} {services collected in 105 days}~\cite{Burda19}. 
Authors in \cite{wang2023comprehensive} {use edit distances to detect phishing and mirror websites, finding the latter in the links of the original HS or in those so-called ``yellow pages'' services.}

However, their contribution differs from our research.  
First, the way they account for mirrors (cryptographic hashes, screenshots\extended{, or edit distances}) can not effectively capture mirror modifications as we do (cf. Figure~\ref{fig:modifications}). 
In particular, they rely on hashes of the HTML content and hashes of the screenshots of the pages for their detection \hl{using Jaccard  Similarity or Hamming distance, which have been proven to not be as effective as our approach in our benchmark on} \S\ref{mirror_analysis}.
\hl{For instance, the latest work is DarkBERT}~\cite{darkbert}
.\hl{ Authors conducted mirror detection using MinHash, achieving a duplication rate of 18.69\%}~\cite{minhash}.
\hl{However, they do not specify the similarity threshold or how they fine-tuned the algorithm. 
Furthermore, since it relies on text tokenization into sets, redundancy in HTML might introduce noise, reducing its effectiveness when using Jaccard similarity.}
Second, 
\textbf{they do not {\em measure} the impact mirrors have on the network topology}. 
\extended{This is, they characterize mirrors alone, but they do not study their interconnection with other hidden services.}

\noindent{\bf Our work.}
Overall, our work differs from related work on three main axes. 
First, we offer a novel and systematic mechanism to discover seeds that prove to be more effective in discovering hidden services.
Second, we consider for the first time the role of mirrors in the measurement of the Dark Web. 
Finally, our system provides a more persistent method to crawl Tor services that is more suitable when dealing with volatile content. 
Table~\ref{tab:related} shows a comparison in terms of coverage of our work with existing works in the literature. 
Since our work does not consider URLs, we do not report these figures (both surface and Hidden Services) seen. 
Our seeding, however, led to nearly 50k Hidden Second-Level Domains (25k available). 
This shows that our crawler sees more sites than previous works, even with a shallow-crawling strategy.

%% file: sections/conclusions.tex
\section{Conclusions}\label{sec:conclusions}

In this paper, we presented the first mirrorless and the largest measurement study of the Tor Network to date. 
We developed a crawler that, together with an innovative seeding method, allows us to study nearly 25k hidden services visited over a period of eight months. 
{While only visiting landing pages, we were able to cover a similar or higher number of sites as in previous works digging further.} 
We showed that the topology of the Dark Web is highly fragmented, with thousands of interconnected independent networks (namely, ``bubbles'') of varying sizes, with a small number of networks having a large number of nodes all highly connected. 

We designed a custom algorithm to detect mirrors and showed that a large proportion of \hl{our dataset} (83\% according to our estimates) consists of replicated content.
Together with an analysis categorizing the content of the different hidden services, we showed the importance of taking into account mirrors when analyzing the Dark Web. 
We found that a large proportion of the network ({67.33\% unique sites and} 74.4\% counting also mirrors) was related to cybercrime activities. Also, we showed the benefits of our methodology to law enforcement and cybercrime researchers through a case study analyzing sites that potentially distribute child abuse material.
Overall, our study offers new insights into the content and structure of Tor, highlighting the importance of using tools that account for duplicate content and paving the way for future research in this area.

\section*{Acknowledgments}

This project received primary funding from TED2021-132900A-I00 and TED2021-132170A-I00, provided by the Spanish Ministry of Science and Innovation through MCIN/AEI /10.13039/501100011033, and the European Union through NextGenerationEU/PRTR. Additional support was provided by PID2022-143304OB-I00, funded by MCIN/AEI /10.13039/501100011033/ and the ERDF "A way of making Europe". Guillermo Suarez-Tangil was appointed as a 2019 Ramon y Cajal fellow (RYC-2020-029401-I), also funded by MCIN/AEI/10.13039/501100011033 and ESF Investing in your future.